 \documentclass[12pt,preprint]{aastex}

\usepackage{graphicx}
\usepackage{natbib}

\def \solphys {Solar Phys.}

\def \aap {A\&A}

\newcommand{\citeN}[1]{\citeauthor{#1} (\citeyear{#1})}
\newcommand{\citeNP}[1]{\citeauthor{#1} \citeyear{#1}}

\shortauthors{Socas-Navarro et al}
\shorttitle{Solar Site Survey for the ATST}

\begin{document}

\title{Solar Site Survey for the Advanced Technology Solar
  Telecope. I. Analysis of the Seeing Data}

\author{H. Socas-Navarro}
   	\affil{High Altitude Observatory, NCAR\thanks{The National Center
	for Atmospheric Research (NCAR) is sponsored by the National Science
	Foundation.}, 3450 Mitchell Lane, Boulder, CO 80307-3000, USA}
	\email{navarro@ucar.edu}
\author{J. Beckers}
       \affil{University of Chicago, IL 60637, USA}
\author{P. Brandt}
       \affil{Kiepenheuer-Institut f\"ur Sonnerphysik, Sch\"oneckstrasse 6,
   	Freiburg, 79104, Germany}
\author{J. Briggs}
       \affil{National Solar Observatory, Sunspot, NM 88349, USA}
\author{T. Brown}
   	\affil{High Altitude Observatory, NCAR, 3450 Mitchell Lane, Boulder, CO
   	80307-3000, USA}
\author{W. Brown}
   	\affil{Atmospheric Technology Division, NCAR, 1850 Table Mesa Dr,
   	Boulder, CO 80307-3000, USA}
\author{M. Collados}
        \affil{Instituto de Astrof\' \i sica de Canarias, Avda V\' \i a L\'
   	actea s/n, E38200, La Laguna (Tenerife), Spain}
\author{C. Denker}
        \affil{New Jersey Institute of Technology, Dr. Martin Luther King
   	Blvd., Newark, NJ 07102-1982, USA}
\author{S. Fletcher}
       \affil{National Solar Observatory, Sunspot, NM 88349, USA}
\author{S. Hegwer}
       \affil{National Solar Observatory, Sunspot, NM 88349, USA}
\author{F. Hill}
       \affil{National Solar Observatory, PO Box 26732, Tucson, AZ
   	85726-6732, USA}
\author{T. Horst}
   	\affil{Atmospheric Technology Division, NCAR, 1850 Table Mesa Dr,
   	Boulder, CO 80307-3000, USA}
\author{M. Komsa}
       \affil{National Solar Observatory, Sunspot, NM 88349, USA}
\author{J. Kuhn}
       \affil{University of Hawaii, Institute of Astronomy, 2680 Woodlawn
   	Drive, Honolulu, HI 96822, USA}
\author{A. Lecinski}
   	\affil{High Altitude Observatory, NCAR, 3450 Mitchell Lane, Boulder, CO
   	80307-3000, USA}
\author{H. Lin}
       \affil{University of Hawaii, Institute of Astronomy, 2680 Woodlawn
   	Drive, Honolulu, HI 96822, USA}
\author{S. Oncley}
   	\affil{Atmospheric Technology Division, NCAR, 1850 Table Mesa Dr,
   	Boulder, CO 80307-3000, USA}
\author{M. Penn}
       \affil{National Solar Observatory, PO Box 26732, Tucson, AZ
   	85726-6732, USA}
\author{T. Rimmele}
       \affil{National Solar Observatory, Sunspot, NM 88349, USA}
\author{K. Streander}
   	\affil{High Altitude Observatory, NCAR, 3450 Mitchell Lane, Boulder, CO
   	80307-3000, USA}

\date{}%

\begin{abstract}
The site survey for the Advanced Technology Solar Telescope concluded
recently after more than two years of data gathering and analysis. Six
locations, including lake, island and continental sites, were thoroughly
probed for image quality and sky brightness. The present paper describes the
analysis methodology employed to determine the height stratification of the
atmospheric turbulence. This information is crucial because day-time seeing
is often very different between the actual telescope aperture ($\sim$30~m)
and the ground. Two independent inversion codes have been developed to
analyze simultaneously data from a scintillometer array and a solar
differential image monitor. We show here the results of applying them to a
sample subset of data from May 2003, which was used for testing. Both codes
retrieve a similar seeing stratification through the height range of
interest. A quantitative comparison between our analysis procedure and actual
{\it in situ} measurements confirms the validity of the inversions. The
sample data presented in this paper reveal a qualitatively different behavior
for the lake sites (dominated by high-altitude seeing) and the rest
(dominated by near-ground turbulence).
\end{abstract}
   

\section{Introduction}
\label{intro}

The Advanced Technology Solar Telescope (ATST, \citeNP{KRK+03}), with its
4-meter aperture, will be the largest solar telescope in the world. A suite
of advanced instrumentation will allow for unprecedented resolution and
precision imaging, spectroscopy and polarimetry at visible and infrared
wavelengths.  In order to address the most challenging problems in modern
solar physics, ATST's highest priorities are oriented towards high-resolution
and coronal science. Obviously, these are aspects in which the site selection
will have a strong impact and therefore should be considered very carefully
in the decision of the telescope location. The ATST Site Survey Working Group
(SSWG) was created in 2000 with an open call to the international
community. Its goal was essentially to test the best solar observatories in
the world and produce a recommendation to maximize the scientific
capabilities of the ATST. More especifically, a set of quantitative goals
were set for the ATST site:
\begin{itemize}
\item Clear time of 70\%, or 3000 annual hours of sunshine.
\item 1800 annual hours of seeing with a Fried parameter $r_0$ better than 7
  cm, including at least 100 continuous 2-hour blocks.
\item 200 annual hours of  $r_0 > 12$~cm, including at least 10 continuous
  2-hour blocks.
\item Large isoplanatic angle for adequate performance of the adaptive optics
  system over a large field of view. This goal is equivalent to demanding
  minimal turbulence at high altitude.
\item 480 annual hours with a sky brightness smaller than 25 millionths of
  the disk value at a distance of 1.1 solar radii. The radial profile curve
  should follow a power law with an exponent $R=0.8$ or steeper. These
  conditions should be met in at least 40 continuous 4-hour block per year.
\item 600 annual hours with precipitable water vapor below 5~mm, including at
  least 40 continuous 4-hour blocks.
\end{itemize}

Only two comparative studies of solar observatories had been conducted prior
to this work. The JOSO organization studied nearly 40 sites in Europe to
conclude that Observatorio del Teide in Tenerife (Canary Islands)
provided the better conditions recorded in their data (\citeNP{BW82};
\citeNP{BR85}). The 
other work is that of the Caltech survey, which considered 34 sites
in the southern California area to locate what is now the Big Bear Solar
Observatory (BBSO; \citeNP{ZM88}).

Initially, a list of 72 well-established observatories and some promising
undeveloped sites was compiled. Since the ATST SSWG had only sufficient
resources for extensive testing of up to six sites, a first decision had to
be made in the absence of any actual measurements. This decision was made
based on previous observing experience of the group members. A first round of
discussion quickly resulted in the selection of BBSO, La Palma (Canary
Islands) and Sacramento Peak (New Mexico), which have an outstanding record
in producing high-quality solar observations. The SSWG also selected Hawaii,
and in particular Haleakala (on the island of Maui) was preferred over Mauna
Loa and Mauna Kea after taking feasibility issues into account. Diverse other
considerations later resulted in the choice of Panguitch Lake (Utah) and San
Pedro M\' artir (Mexico) for the final two slots.

The instrumentation deployed at the six candidate sites includes a
scintillometer array (SHABAR)
and a Solar Differential
Image Motion Monitor (S-DIMM). Both instruments have been described in a
recent paper by \citeN{B01}.
The SHABAR and S-DIMM are used
to determine the atmospheric turbulence as a function of height. It must be
noted that daytime seeing conditions are often strongly dependent on height
above the ground. Thus, the ATST site survey needed to determine the observing
conditions at the telescope aperture height, approximately 35~meters above
the ground (the actual height is still to be decided). The site survey
instrumentation was placed at a height of 8~meters (6-meter stand plus a
2-meter telescope pier), which is high
enough to avoid the lower part of the surface boundary layer but not so high
that wind jitter could become a significant problem.

Data from the SHABAR and S-DIMM instruments have been gathered and analyzed
until August 2004. In October 2004, the ATST SSWG presented a final
recommendation and produced an exhaustive document with a detailed report of
the site survey process. The SSWG report is publicly available for download
from the ATST website (http://atst.nso.edu/). That document contains abundant
technical documentation and the reader is referred to it for further
details. The present series of papers is a formal publication of the most
relevant results. This first paper describes the inversion methods used to
infer the atmospheric turbulence and the Fried parameter from the
SHABAR/S-DIMM data. The methods have been extensively tested by means of
simulations, comparison between different algorithms and also with {\it in
situ} measurements. One month of data (May 2003) was used for testing and
fine tuning the algorithms before being applied systematically to the entire
site survey period.

\section{Derivation of {\bf $C_n(h)$} }
\subsection{Basic relations}
\label{sec:theory}

The final objective is to determine the Fried parameter as a function of
height, $r_0(h)$. This parameter depends not only on the turbulence at a
height $h$ but also everything above it. With some assumptions (see
\citeNP{HL04} for a complete derivation), $r_0$ can be obtained as:
\begin{equation}
\label{eqr0}
r_0^{-5/3} (h)=C \sec (z) \int_h^{\infty} C_n^2(h) dh \, ,
\end{equation}
where $C$ is a constant, $z$ is the zenith angle and $C_n^2$ (which
represents the amplitude of the refraction index fluctuations) characterizes
the local atmospheric turbulence at height $h$. Notice that $h$ denotes
height above the instrument, which in our case is located 8 meters above the
ground.

\begin{figure*}
\plotone{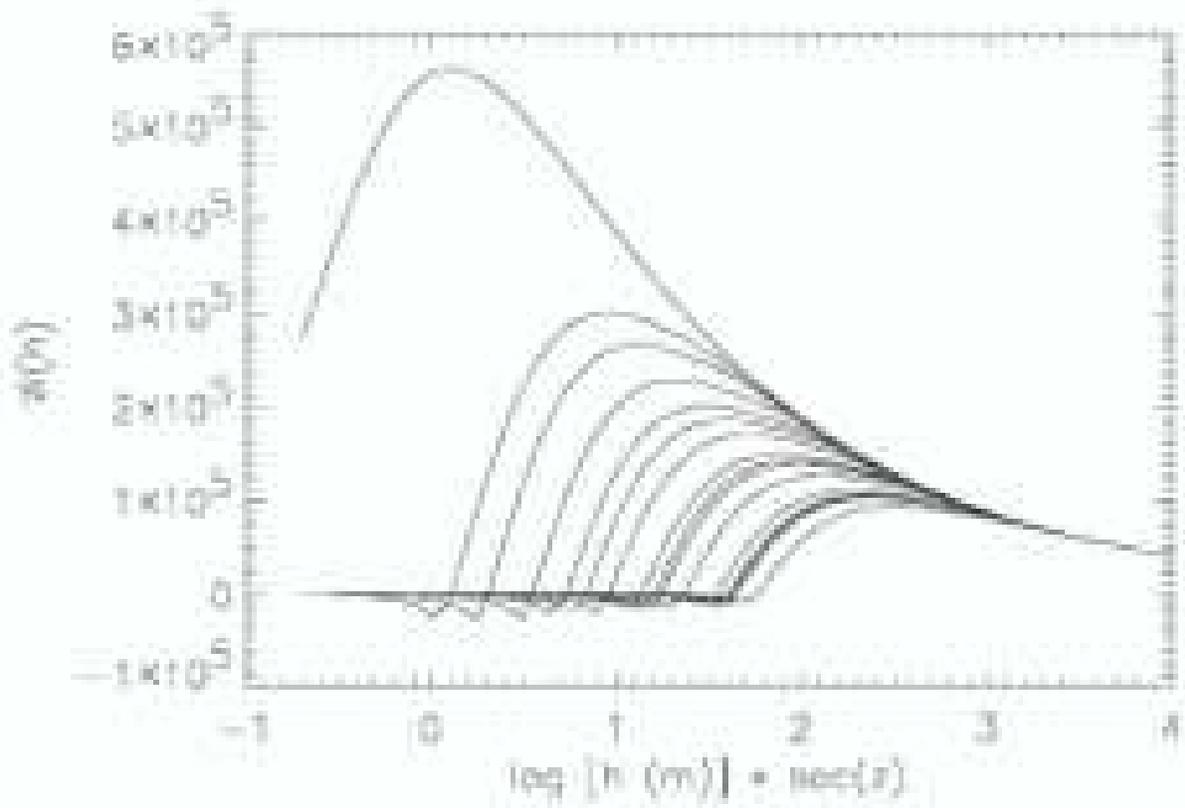}
\caption{
Height dependence of the inversion kernels $W(h,d)$ for the scintillation and
the fifteen covariances (for various values of $d$) between brightness
fluctuations measured by the SHABAR. The upper and lower curves correspond,
respectively, to the smallest and largest sperations (0 and 468~mm). $z$ is
the solar zenith angle.
\label{fig:kernels}
}
\end{figure*}

\begin{figure*}
\plotone{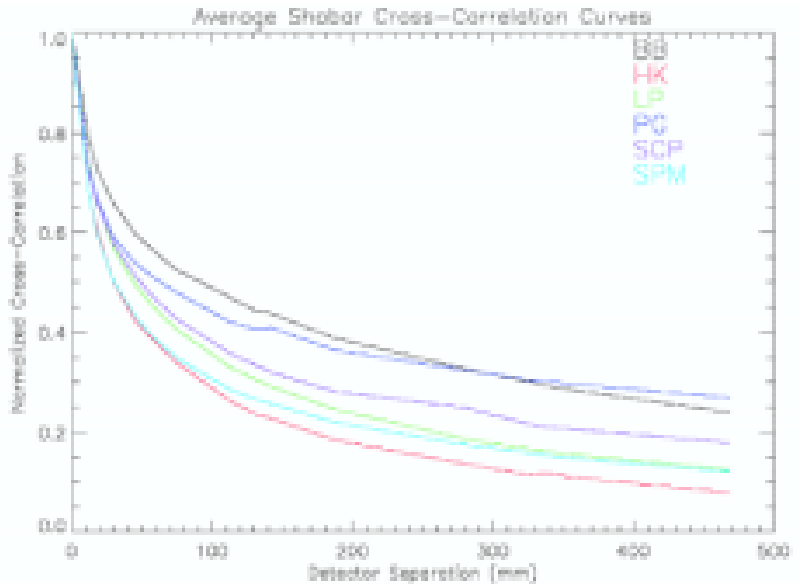}
\caption{
The average cross-correlation curves for the sites as of October 2003. The
two lake sites, Big Bear (BB) and Panguitch (PG), have relatively shallow
curves as a function of detector separation which indicates that the seeing
arises from a region far above the instrument. The four non-lake sites have a
relatively steeper cross-correlation as a function of separation, suggesting
a boundary layer near the telescope.
\label{fig:BIs}
}
\end{figure*}

The covariances ($B_I$) of the brigthness fluctuations measured by two
detectors, separated by a distance $d$ in the SHABAR array, are related to
$C_n^2$ by:
\begin{equation}
\label{eqBi}
B_I(d)= 0.38 \int_0^{\infty} W(h,d)C_n^2(h)dh \, ,
\end{equation}
where the $W(h,d)$ are known functions with analytical expressions.
If $a$ is the detector diameter and $\alpha$ is twice the
tangent of the solar radius (which we consider approximately constant though
the year):
\begin{eqnarray}
W(h,d) &=& {32 \pi h^2 \sec^3 z \over (a + \alpha h \sec z)^{7/3} }
            Q \left ( { d \over a + \alpha h \sec z } \right ) \, ,
            \nonumber \\ 
Q(s) &=& \int_0^{\infty}[J_1(\pi f)]^2 J_0(2\pi f s) f^{-2/3} df \, ,
\end{eqnarray}
with $J_0$ and $J_1$ being the Bessel functions. The total scintillation can
be obtained from the expressions above as $B_I(d=0)$. The kernels $W(h,d)$
are shown in Fig~\ref{fig:kernels}. It can be deduced from this figure that
the $B_I$ corresponding to further separations (larger values of $d$) have
their sensitivity peak at higher atmospheric layers. Fig~\ref{fig:BIs} shows
average values of $B_I(d)$ measured at the six candidate sites. Not
surprisingly, the covariances decrease with the detector separation $d$.

\subsection{The inversion codes}
\label{sec:inv}

The inversion strategy that we employed to solve for $C_n^2(h)$ in
Eq~(\ref{eqBi}) is a non-linear least-squares iterative fitting combined with
singular-value decomposition (SVD) of the covariance matrix. The details of
the implementation can be found, e.g. in \citeN{PFT86}, but the basic idea is
the following. One starts with an initial guess for the $C_n^2(h_i)$ (where
$h_i$ are the gridpoints of the atmospheric discretization). With this
initial model we calculate the 15 different covariances $B_I(d_{jk})$ (where
$d_{jk}$ is the separation between each pair of detectors $j$ and $k$ in the
SHABAR instrument), the total scintillation $B_I(d=0)$ and the Fried
parameter at the base of the atmosphere $r_0(h=0)$ (measured by the S-DIMM
instrument), using Eqs~(\ref{eqr0})
and~(\ref{eqBi}). The derivatives of these parameters with respect to the
$C_n^2(h)$ (the so-called response functions) are also computed, as this
information is needed by the fitting procedure. The synthetic $B_I$ and $r_0$
are then compared to those actually measured by the SHABAR and S-DIMM. In
general, the synthetic values will not match the measurements. The algorithm
then uses the information in the response functions to calculate the
first-order corrections $\delta C_n^2(h_i)$ that minimize the difference (in
a least-squares sense) between the synthetic and measured values of $B_I(d)$
and $r_0(h=0)$.

The numerical solution of an inverse problem such as the one described here
is always challenging, but even more is to ensure that the results obtained
are physically meaningful. Understanding the issues and uncertainties
involved in the inversion is critical for the success of the site survey and,
ultimately, the ATST. It is equally important to ensure that the numerical
codes employed are exhaustively tested and debugged. Based on previous
experience with conceptually similar problems, the SSWG decided that two
different codes should be independently developed and applied to the
data. The two codes, based on the same minimization scheme described above,
exhibit some differences in the actual implementation. By comparing the
results provided by two different codes, we can learn about the reliability
of the procedure and the limitations of our seeing measurements. Since there
is no reason to prefer one above the other, we take the
difference between their respective results as an estimate of the error bar.

\subsubsection{The IAC code}
\label{sec:IAC}

The first code was developed at the Instituto de Astrof\' \i sica de Canarias
(IAC) and run by National Solar Observatory (NSO) staff. The spatial grid is
equispaced in $\log(h)$ with 68 points ranging from 20~cm to 40~km. However,
not all of these points are inverted for (the problem would be ill-posed due
to the large number of free parameters). Instead, $C_n^2(h)$ is determined at
a subset of 17 special locations, the so-called inversion nodes. The rest of
the atmosphere is then reconstructed via splines interpolation from the
values at the nodes.

In order to ensure that $C_n^2(h)$ is always positive, we used the following
change of variable. The code solves for a variable $y$ such that $y(h) = \log
[C_n^2(h)]$. In this manner, regardless of whether $y$ is positive or
negative, $C_n^2(h)$ will always be positive. Before inverting, the measured
$B_I$ and $r_0(h=0)$ are processed with a 5-minute running median filter to
smooth out fluctuations due to instrumental issues or problems such as birds
or insects flying by the detectors.

After an initial application of the code to a sample dataset, it was found
that a significant number of points could not be successfully fitted by the
procedure. The observed $r_0(h=0)$ provided by the S-DIMM instrument was
inconsistent with the observed scintillation values $B_I(d)$ from the
SHABAR. In all cases the S-DIMM $r_0$ was smaller than would be expected from
the scintillation measurements. To overcome this difficulty we introduced an
additional parameter $\Delta s$ in the inversion that accounts for this
``missing scintillation'', so that the synthetic $B_I$ are now given by:
\begin{equation}
\label{eqBimcv}
B_I(d) + \Delta s = 0.38 \int_0^{\infty} W(h,d)C_n^2(h)dh  \, ,
\end{equation}
while Eq~(\ref{eqr0}) is still used to compute $r_0(h)$. 

\begin{figure*}
\plotone{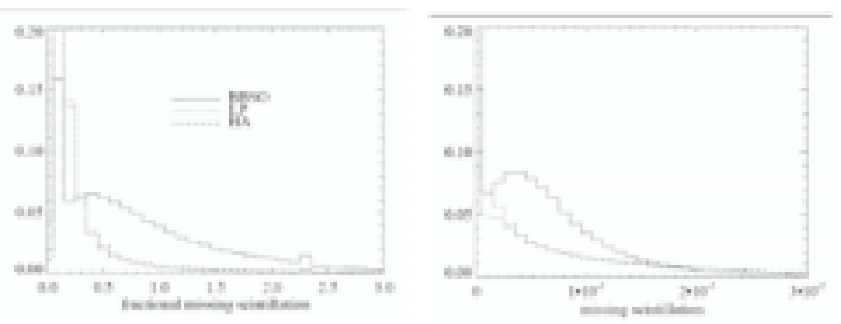}
\caption{
Histograms of the relative frequency of occurrence of missing scintillation
$\Delta s$ values at the three sites. 
Left: relative occurrence of
$\Delta s$ as a fraction of the observed scintillation; left: entire
distribution; 
Right: relative occurrence of $\Delta s$ as an absolute scintillation
measure. Solid line: 
Big Bear; Dashed line: Haleakala; Dotted line: La Palma. Note that the two
mountain sites (La Palma and Haleakala) typically require 10-20\% fractional
$\Delta s$ while the lake site (Big Bear) frequently needs a substantially
larger value.
\label{fig:deltas}
}
\end{figure*}

The necessity to include $\Delta s$ points towards the existence of a a
source of image degradation that does not produce scintillation. This may be
due to a finite outer turbulence scale, to ``windfiltering'' (due to the
finite time amount it takes for the turbulence to cross the field-of-view at
high altitudes) or to high-altitude seeing (higher
than $\sim$1~km, where the kernels converge). In either case, the inclusion
of $\Delta s$ would be appropriate. While we have not been able to pinpoint
the exact nature of the missing scintillation, the statistical distribution
of $\Delta s$ (see Fig~\ref{fig:deltas}) provides an interesting clue.

The two island sites (La Palma and Haleakala) require a relatively small
amount of missing scintillation, whereas the lake site (Big Bear) frequently
demands a substantially larger value. As we discuss below, the seeing in the
island sites is typically dominated by a boundary layer near the ground and
the turbulence decreases rapidly as we move upwards. The lake sites, on the
other hand, are frequently dominated by high-altitude seeing. Thus, it seems
likely that the $\Delta s$ is related to the turbulence above the sensitivity
range of the SHABAR.

\subsubsection{The HAO code}
\label{sec:HAO}

The second inversion code was developed independently at the High Altitude
Observatory (HAO). One of the essential differences with the IAC code is that
all the grid-points in the atmospheric discretization are now free
parameters. In order to make the algorithm stable, a regularization scheme was
implemented in such a way that the code prefers smooth solutions whenever
possible. Basically, in each iteration we add a penalization term to the
$\chi^2$. This term is proportional to the quadratic deviation of 
$C_n^2(h)$ from a straight line on a $\log$-$\log$ plot. The fact that each
grid-point is a free parameter represents an important burden that slows down
the execution of the code. This is compensated, to some extent, by the
more accurate integration scheme described below.

The model atmosphere only goes up to a maximum height $H_m = 1000/\sec(z)$
where the inversion kernels merge (Fig~\ref{fig:kernels}). We still consider
the effect of turbulence above $H_m$ by breaking down the integrals as:
\begin{equation}
\label{eqBi2}
B_I(d)= 0.38 \left ( \int_0^{H_m} W(h,d)C_n^2(h)dh + B_I^{High} \right ) \,
\end{equation}
and
\begin{equation}
\label{eqr02}
r_0^{-5/3} (h)=C \sec (z) \left ( \int_h^{H_m} C_n^2(h) dh +
   \alpha B_I^{High} \right ) \, .
\end{equation}
Note that $B_I^{High}$ does not depend on $d$.  As before, $\alpha$ and
$B_I^{High}$ are free parameters that account for the scintillation and image
degradation coming from high-altitude turbulence.

Another important difference with the IAC code is the integration scheme used
to solve Eqs~(\ref{eqBi2}) and~(\ref{eqr02}). The integral over each finite
grid interval is computed assuming a parabolic dependence of $\log(C_n^2)$
with $\log(h)$ (see the appendix for details). While this integration
introduces some additional complexity in the algorithm, it also alleviates
the computational expense of the inversions by allowing a coarser grid
without loss of accuracy. 

The data pre-processing includes a 5-minute block average and a test to look
for clouds in the averaged interval. If the derivative of the intensity
measured by the SHABAR detectors changes sign two or more times, the entire
period is flagged as cloudy and the $r_0(h)$ is set to zero at all heights.

\section{Verification tests}
\label{sec:tests}

\subsection{Comparison of two codes}
\label{sec:codes}

\begin{figure*}
\plotone{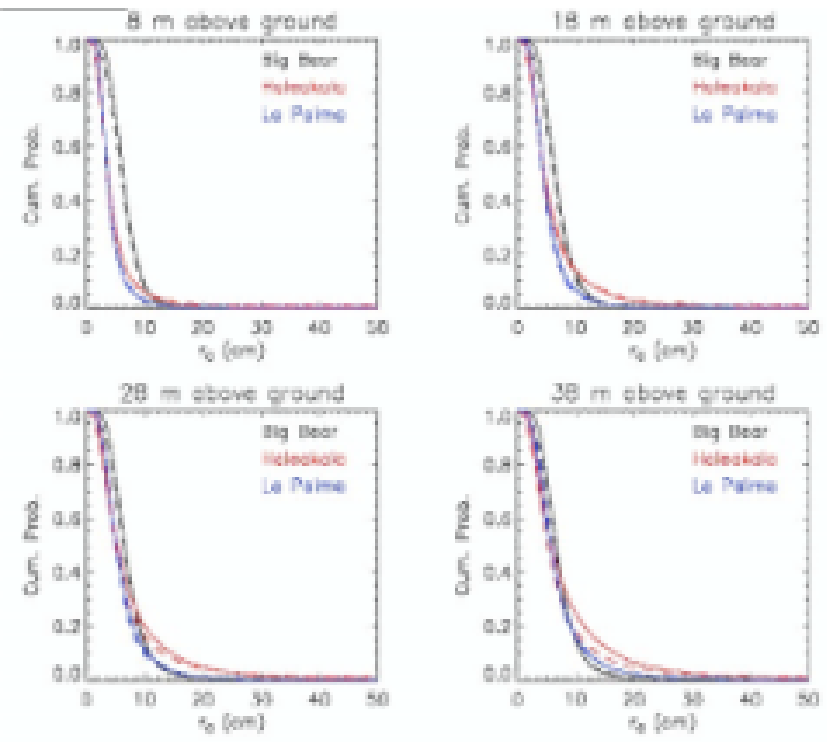}
\caption{
A comparison between the two inversion methods. Here we show the cumulative
distribution of the esimated $r_0(h)$ at the three sites and at four heights,
as derived from the HAO method (dashed line) and the IAC method (solid line).
\label{fig:testcomp1}
}
\end{figure*}

A sample subset of data (those taken in May 2003) were subject to inversion
with both codes. Fig~\ref{fig:testcomp1} shows the cumulative distributions
produced by each code at La Palma, Haleakala and Big Bear. The results are
fairly similar for most of the height range represented in the figure. As one
would expect, the discrepancy increases with height due to the limited
sensitivity of the SHABAR measurements to turbulence at higher layers. The
same behavior is observed with the in situ measurements (see
section~\ref{sec:insitu} below).

\begin{figure*}
\plotone{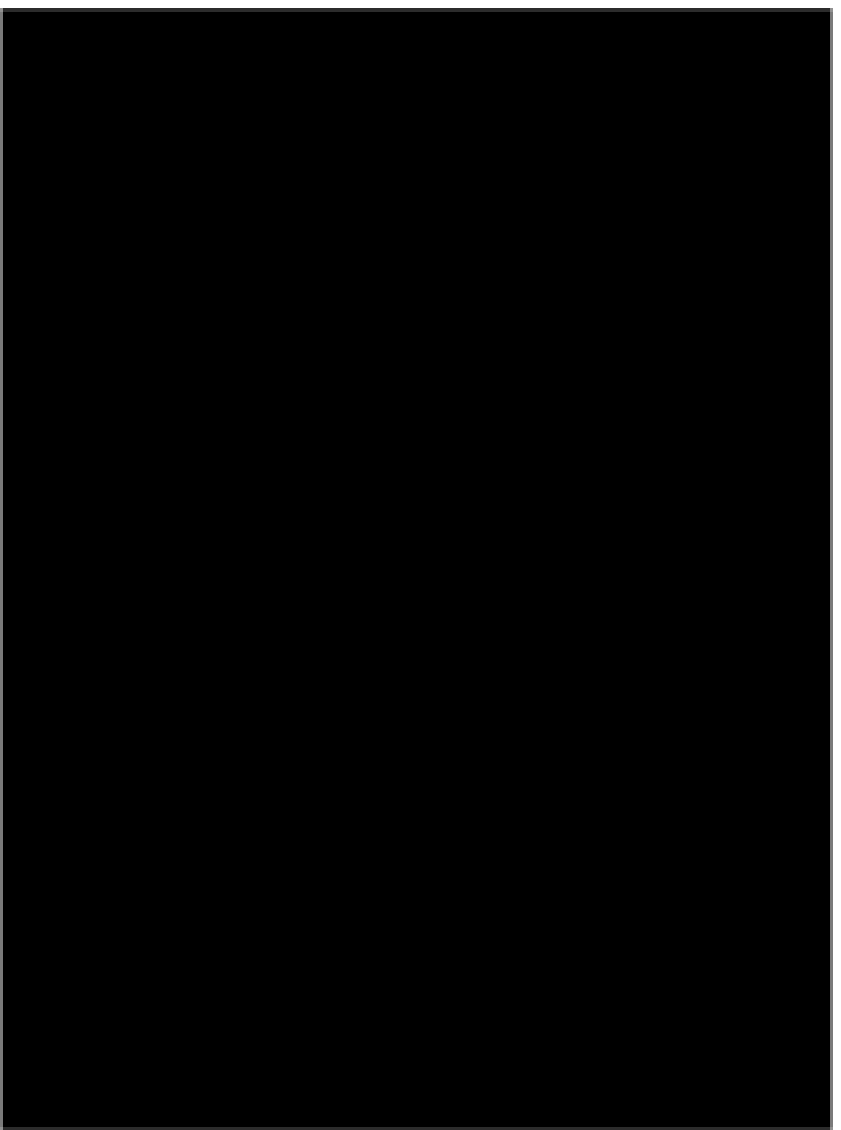}
\caption{
Curves of $C_n^2(h)$ and $r_0(h)$ retrieved by the inversions averaged over
May 2003 for Big Bear, Haleakala and La Palma. Dashed: HAO method, Solid: IAC
method. 
\label{fig:testcomp2}
}
\end{figure*}

\begin{figure*}
\plotone{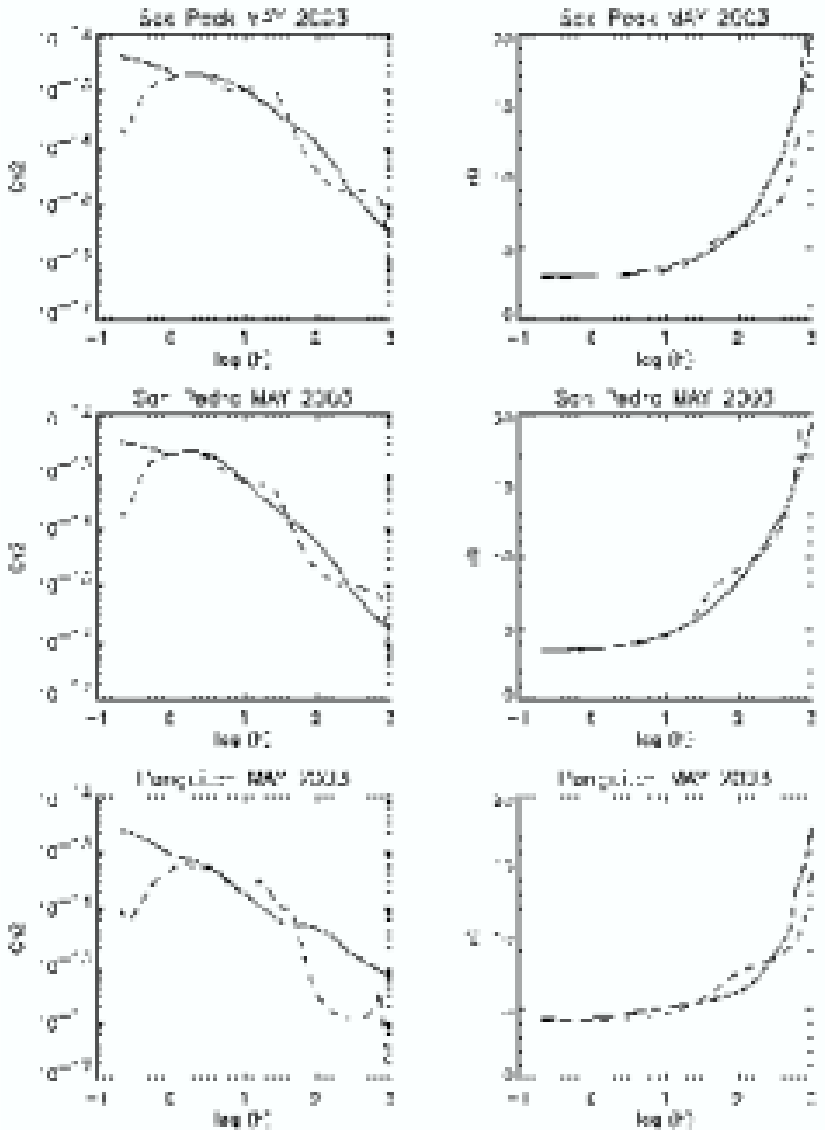}
\caption{
Curves of $C_n^2(h)$ and $r_0(h)$ retrieved by the inversions averaged over
May 2003 for Sac Peak, San Pedro M\' artir and Panguitch. Dashed: HAO method,
Solid: IAC method.
\label{fig:testcomp3}
}
\end{figure*}

A more detailed representation is given in Figs~\ref{fig:testcomp2}
and~\ref{fig:testcomp3}, which show the average stratifications of $C_n^2$
and $r_0$ with height. The $C_n^2$ produced by the HAO code has more spatial
structure (small-scale fluctuations) than the IAC code. However, the overall
dependence is actually very similar and the IAC results resemble a
smoothed-out version of those from HAO (at least down to the first meter
above the detectors). Whether the fine structure inferred by the HAO code is
real or not is practically irrelevant for our purposes here, as the $r_0(h)$
parameter, which is our ultimate goal, is barely sensitive to it. The $r_0$
stratifications from both codes are very similar, at least up to $\sim$~40~m
above the instrument. In this small sample we can already see the different
behavior of the lake sites (Big Bear and Panguitch) as opposed to the island
(La Palma and Haleakala) and continental sites (San Pedro M\' artir and Sac
Peak). The Big Bear plot exhibits a larger $r_0$ near the ground, but it does
not improve significantly with height. The island and continental sites, on
the other hand, have a lower value at the ground but the seeing improves
dramatically as we go to higher layers, suggesting the presence of a boundary
layer near the ground. The slope is particularly steep in the $\sim$30-40~m
range, precisely where the ATST aperture will be. These results emphasize the
importance of analyzing the height dependence of $r_0$, as we have done in
this work, since the site and height of the telescope may end up being
critical factors in the performance of the ATST.

\subsection{In situ measurements}
\label{sec:insitu}

Perhaps the most reliable test of the ATST site survey analysis is to compare
its results directly with actual in situ measurements. The SSWG had access to
a tower with meteorological instrumentation including hygrometers and sonic
anemometers at various heights (DASH, \citeNP{OH04}). With this
instrumentation it is possible to make local measurements of temperature $T$
and humidity $q$ up to a height of $\sim$100~m. These parameters are
monitored at a rate of 30~Hz, from which $C_n^2$ can be derived.

\begin{figure*}
\plotone{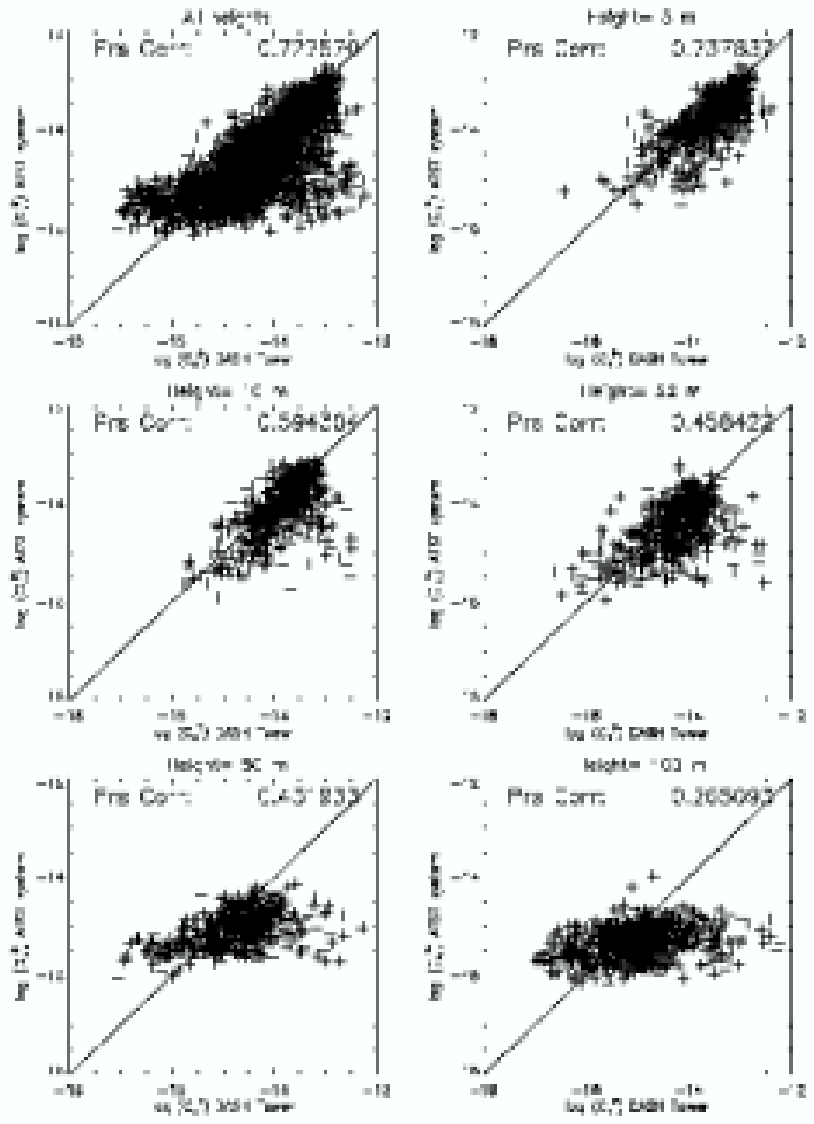}
\caption{
Comparison of $C_n^2$ from the ATST system with in situ measurements in Erie,
Colorado. 
\label{fig:erie}
}
\end{figure*}

\begin{figure*}
\plotone{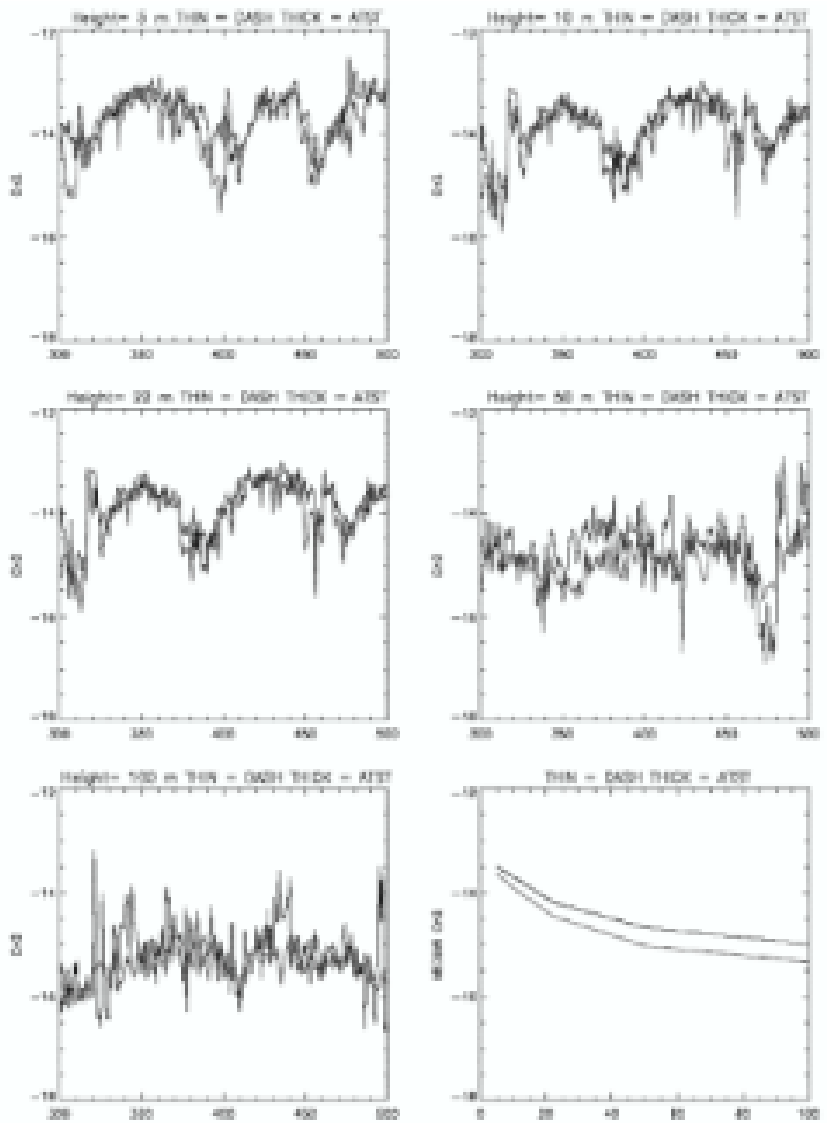}
\caption{
Time series determinations of $C_n^2$ with the Erie tower and the ATST
instrumentation at 5, 10, 22, 50 and 100~m. Bottom-right panel: Median value
as a function of height.
\label{fig:erie2}
}
\end{figure*}

A complete SHABAR/S-DIMM system was deployed in Erie (Colorado), at the base
of the measuring tower. Simultaneous measurements from the tower and the ATST
instrumentation have been used to produce the scatter plots in
Fig~\ref{fig:erie}. Each panel shows the diagonal of the plot, as well as the
Pearson correlation between the ATST inversions and the SHABAR
instrumentation. Before analyzing the discrepancies between the two, a brief
explanation is in order. While the tower measurements are strictly local, the
SHABAR detectors receive solar light that has passed through a cone of Earth
atmosphere. Thus, the ATST instrumentation is sensitive to a field of view
that becomes larger for higher heights. For this reason, the $C_n^2$ measured
in situ is expected to exhibit larger temporal fluctuations than that from
the ATST instrumentation, which is a spatial average over the SHABAR field of
view. This effect, which can be seen in Fig~\ref{fig:erie2}, increases the
scatter especially in the higher layers. Moreover, the SHABAR field of view
is displaced with respect to the tower instrumentation by an amount that
increases with height. In any case, Fig~\ref{fig:erie2} shows that the
overall trends are in general well retrieved by the ATST analysis.

The tests indicate good agreement between both types of measurements up to a
height between 22 and 50~m. It is important to note that the height range
where the ATST measurements are accurate is lower in this case than at the
actual candidate sites, simply because the tests were carried out during
mid-winter at the considerably higher latitude of Erie (the $\sec[z]$
angle was always larger than 2.0).

\begin{figure*}
\plotone{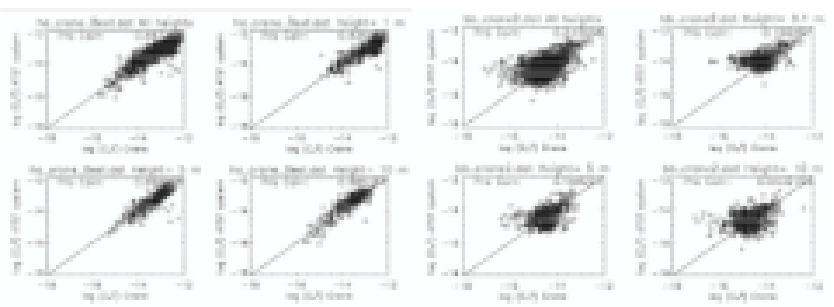}
\caption{
Scatter plot comparisons between the ATST estimates and the in situ
measurements of $C_n^2$ at Haleakala (left four panels) and Big Bear (right
four panels). In each four-panel set the comparison is for the three heights
of the in situ measurements, plus all of the points combined. The straight
line is strict equality, and each plot is labeled with the Pearson
correlation coefficient of the data.
\label{fig:tests2}
}
\end{figure*}

While the Erie tests reassured our confidence in the ATST seeing
measurements, there was still some concern that the atmospheric structure
above the continental Colorado is probably very different from the actual
sites. The SSWG then decided to mount similar in-situ probes on a portable
crane to perform similar tests on Big Bear and Haleakala.
New measurements (this time considering lower heights due to the
limitations of the crane) were conducted at the two sites, with the results
shown in Fig~\ref{fig:tests2}.

The Haleakala tests show excellent agreement between the inversions and the
measurements. In Big Bear, however, the correlation is very weak. Among the
differences between the two sites are the level of humidity, the local
topography and environment, and the probable lack of a ground boundary layer
in Big Bear.

\begin{figure*}
\plotone{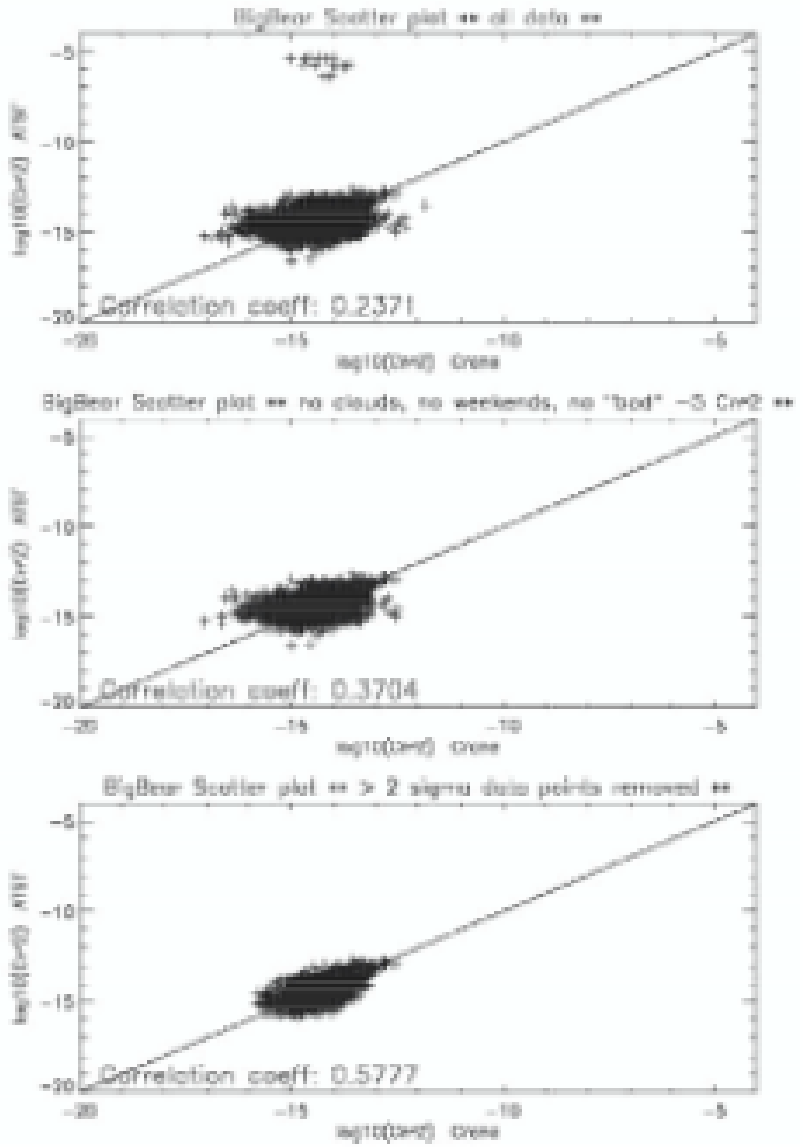}
\caption{
The influence of noise on the correlation between the ATST and in situ
measurements of $C_n^2$. Top: all data. Middle: excluding possibly cloudy
points. Bottom: excluding points more than 2-$\sigma$ from the mean. 
\label{fig:bbtest1}
}
\end{figure*}

\begin{figure*}
\plotone{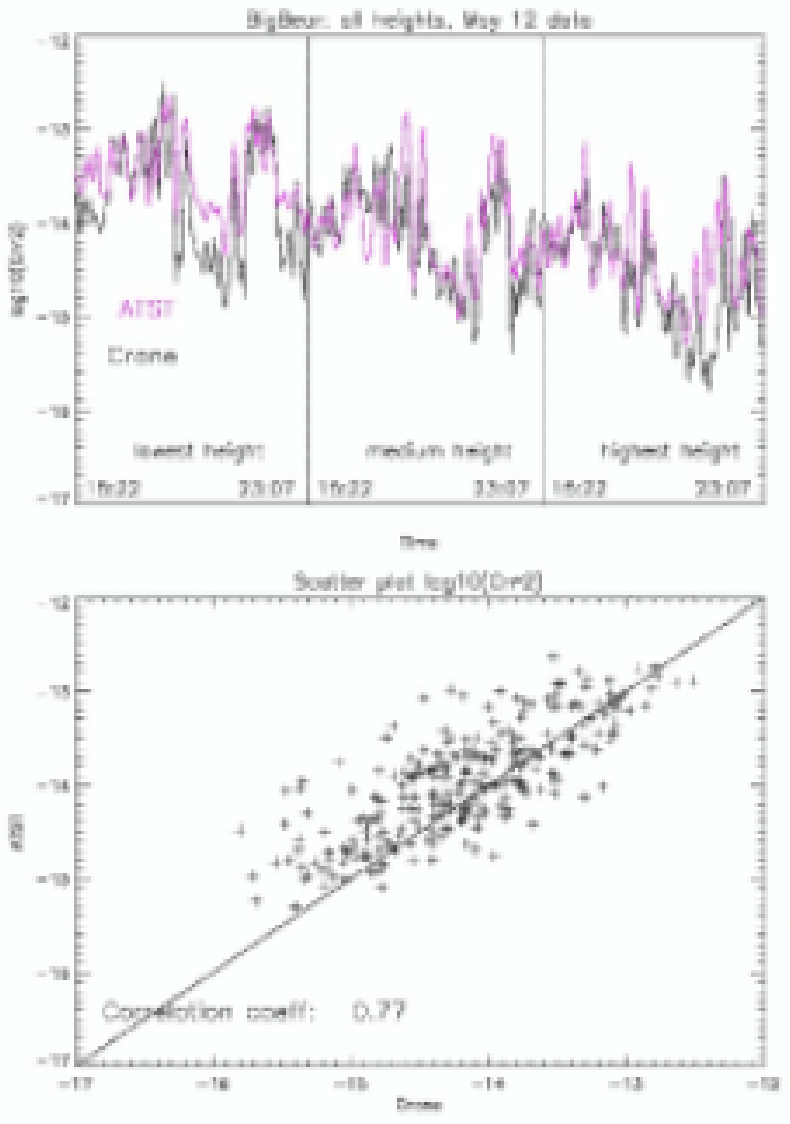}
\caption{
The correlations for Big Bear on May 12, 2004 when the winds were atypically
variable. Top: time series for the three heights. Bottom: correlation for
this day.
\label{fig:bbtest2}
}
\end{figure*}

A possible explanation may be that the weaker scintillation in Big
Bear makes these measurements more sensitive to noise. A closer look at the
data shows that the correlation is improved by removing outlier points and
clouds (Fig~\ref{fig:bbtest1}). Particularly interesting is the dataset
gathered on May 12 2004, which reveals the dependence of the data with the
wind direction (Fig~\ref{fig:bbtest2}). When the wind blows from the lake
(west), the observing conditions are optimal and the seeing is dominated by
high-altitude turbulence. On that particular day the wind changed from west
to variable directions, resulting in the development of a ground turbulence
layer and improving the correlation coefficient to $\sim$0.77. Note also
that, in the absence of a surface turbulence layer at Big Bear, $r_0$ will
essentially be independent of $h$ near the ground and the S-DIMM measurements 
at 8~m can be taken to represent $r_0$ at 25-40~m.


\section{Conclusions}
\label{sec:conc}

The ATST project has imposed stringent requirements on the telescope site,
both in terms of seeing and sky brightness. The seeing analysis has proven
very challenging due to the strong dependence of $r_0$ with height exhibited
by most sites. This paper describes the analysis employed by the ATST SSWG to
find the site that provides optimal conditions ($r_0$) at the height of the
ATST primary aperture. 

Various tests have been used to demonstrate the validity of our analysis, at
least up to $\sim$40~m above the ground. The use of two independent inversion
codes are helpful to understand the uncertainties in the procedure and to set
error bars on the measurements. Moreover, in situ measurements proved the
validity of the approach in the height range of interest. The only caveat
identified by our work is that the errors increase when the seeing is
dominated by high-altitude turbulence (i.e., in the absence of a ground
boundary layer).

\appendix
\section{Parabolic integration scheme}
\label{appendix}

Let $x=\log(h)$ and $y=\log[C_n^2(h)]$. If we approximate $y$ and $W$ as
parabolic functions of $x$:
\begin{eqnarray}
W & = & w_1 x^2 + w_2 x + w_3 \\ \nonumber
y & = & a x^2 + b x + c  \, ,
\end{eqnarray}
we can solve analytically the integral in
Eq~(\ref{eqBi2}) between two adjacent grid-points $x_1$ and $x_2$, as:

\begin{eqnarray}
I&=&\int_{x_1}^{x_2}\left[w_1x^2+w_2x+w_3\right]
   10^{\left(ax^2+bx+c\right)}dx\\ \nonumber
&=&10^c\int_{x_1}^{x_2}\left[w_1x^2+w_2x+w_3\right]
   10^{\left(ax^2+bx\right)}dx\\ \nonumber
&=&10^c\left(T_1+T_2+T_3\right)
\end{eqnarray}
where,
\begin{eqnarray*}
T_1&=&\frac{{2^{-3 - \frac{b^2}{4\,a}}\, {w_1}\,}}
      {5^{\frac{b^2}{4\,a}}\,a^{\frac{5}{2}}\,{\log (10)}^{\frac{3}{2}}}
    \left\{ 2^{1 + \frac{b^2}{4\,a}}\,5^{\frac{b^2}{4\,a}}\,{\sqrt{a}}\,
       \left( \left( {10}^
             { {x_1}\,\left( b + a\, {x_1} \right) } - 
            {10}^{ {x_2}\,\left( b + a\, {x_2} \right) }
          \right) \,b
\right.\right.\\&&\left.\left.
     + 2\,a\,\left( -\left( {10}^
                { {x_1}\,\left( b + a\, {x_1} \right) }\,
                {x_1} \right)  +  
            {10}^{ {x_2}\,\left( b + a\, {x_2} \right) }\,
              {x_2} \right)  \right) \,{\sqrt{\log (10)}}
\right.\\&&\left.
      + {\sqrt{\pi }}\, {\rm Erf}(\frac{\left( b + 
             2\,a\, {x_1} \right) \,{\sqrt{\log (10)}}}{2\,
           {\sqrt{a}}})\,\left( 2\,a - { {b}}^2\,\log (10) \
\right) \right.\\&&\left.
+ {\sqrt{\pi }}\, {\rm Erf}(\frac{\left( b + 
             2\,a\, {x_2} \right) \,{\sqrt{\log (10)}}}{2\,
           {\sqrt{a}}})\,\left( -2\, {a} + 
         { {b}}^2\,\log (10) \right)  \right\} \\
T_2&=& \frac{2^{-2 - \frac{b^2}{4\,a}}\, {w_2}\,}
    {5^{\frac{b^2}{4\,a}}\,a^{\frac{3}{2}}\,\log (10)}
    \left\{ 2^{1 + \frac{b^2}{4\,a}}\,5^{\frac{b^2}{4\,a}}\,
       \left( -{10}^
           { {x_1}\,\left( b + a\, {x_1} \right) } + 
         {10}^{ {x_2}\,\left( b + a\, {x_2} \right) } \
\right) \,{\sqrt{a}} 
       \right.\\&&\left.
       +b\, {\rm Erf}(\frac{\left( b + 
             2\,a\, {x_1} \right) \,{\sqrt{\log (10)}}}{2\,
           {\sqrt{a}}})\,{\sqrt{\pi \,\log (10)}}
       \right.\\&&\left.
      - b\, {\rm Erf}(\frac{\left( b + 2\,a\, {x_2} \right) \,
           {\sqrt{\log (10)}}}{2\,{\sqrt{a}}})\,{\sqrt{\pi \,\log (10)}} \
\right\} \\
T_3&=& -\frac{2^{-1 - \frac{b^2}{4\,a}}\, {w_3}\,}
     {5^{\frac{b^2}{4\,a}}\,{\sqrt{a}}}
      \left({\sqrt{\frac{\pi }{\log (10)}}} \right) 
    \left\{ {\rm Erf}(\frac{\left( b + 2\,a\, {x_1} \
\right) \,{\sqrt{\log (10)}}}{2\,{\sqrt{a}}})
     \right.\\&&\left.
        - {\rm Erf}(\frac{\left( b + 2\,a\, {x_2} \right) \,
            {\sqrt{\log (10)}}}{2\,{\sqrt{a}}}) \, \right\} 
\end{eqnarray*}

\acknowledgments


\begin{thebibliography}{8}
\expandafter\ifx\csname natexlab\endcsname\relax\def\natexlab#1{#1}\fi

\bibitem[{{Beckers}(2001)}]{B01}
{Beckers}, J.~M. 2001, Experimental Astronomy, 12, 1

\bibitem[{{Brandt} \& {Righini}(1985)}]{BR85}
{Brandt}, P.~N., \& {Righini}, A. 1985, Vistas in Astronomy, 28, 437

\bibitem[{{Brandt} \& {Woehl}(1982)}]{BW82}
{Brandt}, P.~N., \& {Woehl}, H. 1982, \aap, 109, 77

\bibitem[{{Hickson} \& {Lanzetta}(2004)}]{HL04}
{Hickson}, P., \& {Lanzetta}, K. 2004, \pasp, 116, 1143

\bibitem[{{Keil} {et~al.}(2003){Keil}, {Rimmele}, {Keller}, {Hill}, {Radick},
  {Oschmann}, {Warner}, {Dalrymple}, {Briggs}, {Hegwer}, \& {Ren}}]{KRK+03}
{Keil}, S.~L., {Rimmele}, T., {Keller}, C.~U., {Hill}, F., {Radick}, R.~R.,
  {Oschmann}, J.~M., {Warner}, M., {Dalrymple}, N.~E., {Briggs}, J., {Hegwer},
  S.~L., \& {Ren}, D. 2003, in Innovative Telescopes and Instrumentation for
  Solar Astrophysics. Edited by Stephen L. Keil, Sergey V. Avakyan .
  Proceedings of the SPIE, Volume 4853, pp. 240-251 (2003)., 240--251

\bibitem[{Oncley \& Horst(2004)}]{OH04}
Oncley, S., \& Horst, T. 2004, Calculation of $C_n^2$ for visible light and
  sound from CSAT3 sonic anemometer measurements, available at
  http://www.atd.ucar.edu/homes/oncley/bao2004/background.pdf

\bibitem[{{Press} {et~al.}(1986){Press}, {Flannery}, \& {Teukolsky}}]{PFT86}
{Press}, W.~H., {Flannery}, B.~P., \& {Teukolsky}, S.~A. 1986, Numerical
  recipes. The art of scientific computing (Cambridge: University Press, 1986)

\bibitem[{{Zirin} \& {Mosher}(1988)}]{ZM88}
{Zirin}, H., \& {Mosher}, J.~M. 1988, \solphys, 115, 183

\end{thebibliography}

\end{document}